\documentclass[journal,onecolumn,draftclsnofoot]{IEEEtran}

\usepackage{cite}
\usepackage{array}
\usepackage{graphicx}
\usepackage{amsmath}
\usepackage{amsfonts}
\usepackage{amstext}
\usepackage{multirow}
\usepackage{amssymb}
\usepackage[english]{babel}

\usepackage[english]{babel}
\usepackage[utf8]{inputenc}
\usepackage{algorithm}
\usepackage[noend]{algpseudocode}

\usepackage{amsthm}

\newtheorem*{remark}{Remark}

\ifCLASSINFOpdf
\else
\fi
\begin{document}%
\title{
An Accurate Fault Location Algorithm for Meshed Power Networks Utilizing Hybrid Sparse Voltage and Current Measurements}
%
%
%

\author{Adil~Khan, Abdul~Qayyum~Khan, Muhammad~Sarwar, Muhammad~Abubakar, Naeem~Iqbal
}

\maketitle

\begin{abstract}
This paper proposes an accurate fault location algorithm technique based on hybrid synchronized sparse voltage and sparse current phasor measurements. The proposed algorithm addresses the performance limitation of fault location algorithms based on only synchronized sparse voltage measurements (SSVM) and on only synchronized sparse current measurements (SSCM). In the proposed method, bus voltage phasor of faulty line or close to faulty line and branch current phasor of the adjacent line is utilized. The paper contributes to improve the accuracy of fault location and deter the effect of CT saturation by using hybrid voltage and current measurements. The proposed algorithm has been tested on four bus two area power system and IEEE 14 bus system with the typical features of an actual distribution system. The robustness of algorithm has been tested by variation in fault location, fault resistance, load switching. The simulation results demonstrate the accuracy of proposed algorithm and ensures a reliable fault detection and location method.

\end{abstract}

%
\IEEEpeerreviewmaketitle

\section{Introduction}
%
%
%
%
\IEEEPARstart{A}{ccurate} and timely fault detection on power lines results in quick restoration of power, speeds up inspection and repair of faulty power transmission line. It reduces customer's complaints, power outage time and loss of revenue; as a result, reliability of power system increases.

Faults are inevitable in power system and mitigation and isolation of faults is of cardinal importance for reliable operation of the power system. Owing to the importance of fault detection and location in power system, many researchers have proposed algorithms for identification and isolation as well as prediction accurate location of the fault.

Global Positioning System (GPS) based Phasor Measurment Units (PMUs) have been deployed for accurate phasor measurements since late 1980s. Consequently, Wide Area Measurement/Monitoring System (WAMS) concept has been developed for upgrading the supervision, protection, operation and control of modern power systems. GPS based PMUs provide synchronized measurements for accurate monitoring and control of the power system in real-time. The same measurements can be used for the fault location and detection algorithms \cite{1r,2r}.



Impedance-based fault location techniques utilize the voltages and currents signals at the fundamental frequency. Voltage and current phasors can be provided by PMUs installed at important buses in the power system. Depending upon the input signals, fault location algorithms are classified into one-end, two-end and multi-ended algorithms. One-end algorithms utilize voltages and currents from only one end of the line for estimating the distance to the fault. Such techniques \cite{3r,4r,5r,6r,7r,8r,29r} are simple and do not require communication with the remote end of the transmission line. Single-ended algorithms suffer from the low accuracy due to higher fault resistance, zero sequence mutual coupling, tapped load and non homogeneous power systems.

Two-end algorithms utilize signals from both ends of the line and thus more amount of information about the power system is utilized. As a result, two-end algorithms have superior performance than one-end algorithms. The input signals to the two-end algorithms are current and voltage measurements \cite{9r,10r,11r,12r,13r,14r,15r,16r,17r}. In \cite{18r}, only current measurements are utilized to develop fault location algorithms. In \cite{19r,20r,21r}, only voltage measurements of both ends are utilized in order to ensure complete immunity to current transformer (CT) saturation. The algorithm proposed in \cite{22r,23r} utilizes minimum amount of information between line terminals.


A fault location algorithm using only Synchronized Sparse Voltage Measurements (SSVM) of both-end was proposed in \cite{24r}. The algorithm utilizes only voltage measurements from buses, which may be away from the faulted line. 

Ali \textit{et al} proposed a fault modeling and detection procedure using analytical techniques to detect and isolate faults in power transmission and distribution networks \cite{ietgtd}.
A fault location algorithm is proposed in \cite{ifac} based on positive sequence voltage and current measurements obtained from PMUs placed at strategic locations in the power system.
Liao \cite{25r} proposed fault location method utilizing only Synchronized Sparse Current Measurements (SSCM) from the branches which may be distant from the faulted section. This algorithms seems more accurate than the voltage based methods because of having more information about the system. However, as the algorithm utilize only current measurements so its performance is limited by the CT saturation due to higher magnitude fault current.

Traditionally, existing two-ended methods [2, 7-9] require the phase alignment of data sets captured at both ends of a monitored line using pre-fault load flow information, iterative methods, and communication of a significant amount of data between relay terminals. Other researchers [7-10] have proposed different methodologies using fundamental frequency phasor data from two terminals of a line, and in some cases from three- or multi-terminal lines. These methods have some inherent limitations, such as requirement for data alignment, knowledge of pre-fault load flow information, need to perform iterations, and communication of a large amount of data between the terminals. In addition, a number of multi-terminal methods [7,9-10] are not applicable to overhead lines with zero-sequence mutual coupling.

To overcome the shortcomings of above mentioned algorithms, this paper proposes a hybrid fault location algorithm utilizing both synchronized sparse voltage and sparse current phasor measurements. The proposed method addresses the performance limitation of fault location algorithms based on only voltage \cite{24r} or only current measurement \cite{25r}. The data required for the estimation of fault location does not require load-flow information and the amount of data transmitted between the terminals is sufficiently small and can easily be transmitted using a single communication channel. Furthermore, the proposed algorithm is non-iterative and does not require information of phase-selection for asymmetrical faults. The algorithm is also immune to variations in fault impedance, power system non-homogeneity and tapped loads.

The new algorithm is rigorously tested for symmetrical and unsymmetrical faults with varying fault impedance on two disparate systems with different parameters. The robustness of the algorithm is tested for a variety of practical system operating conditions and configurations. The results demonstrate an accuracy improvement over the single-ended methods and the algorithms which utilize only voltage or current measurements for fault location estimation.

A comparison is also made between the proposed algorithm and the fault locations method based on only SSVM \cite{24r}, and on only SSCM \cite{25r}. The proposed method is more accurate than the SSVM and also immune to CT saturation. It is assumed that network is transposed and its data is known. The PMUs are optimally placed at a limited number of buses and the fault detection and faulty line is identified with available number of PMUs.

The organization of the rest of the paper is as follows. In Section II, fault location using SSVM is discussed. Section III describes the fault location algorithm utilizing only SSCM. In Section IV, proposed algorithm is discussed. Simulation results to prove the effectiveness of the method are demonstrated in Section V.

\section{Algorithm Based on Synchronized Sparse Voltage Measurements (SSVM)}
\begin{figure}
	\centering
	\includegraphics[width=0.5\textwidth]{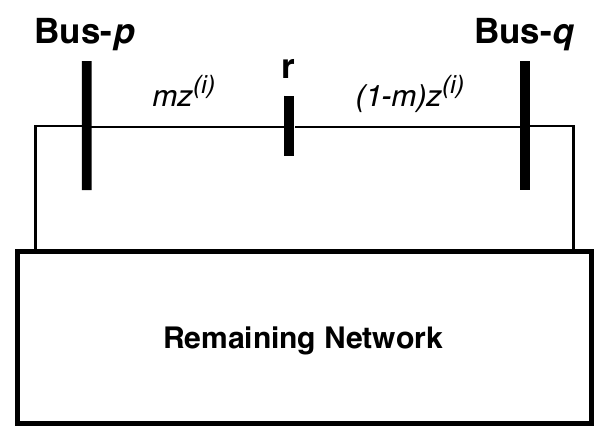}
	\caption{Sample Power System for finding bus impedance matrix for a fault at fictitious bus $r$}
	\label{SAMPLE}
\end{figure}

The proposed algorithm is based on the bus impedance matrix technique \cite{26r}. A fictitious bus is added at the fault point and the impedance of all the buses is expressed as a function of fault location. The fault location is derived based on the relationship between change in bus voltage due to fault and the transfer impedance. A sample power system having $n$ buses has been shown in Fig. \ref{SAMPLE}.

$z^{(i)}$ is the total impedance of the line between bus $p$ and $q$ where $i$ denote sequence quantity. $i = 0, 1 $ and $ 2$ indicate zero, positive and negative sequence quantities respectively. The bus impedance matrix of the pre-fault power system $Z_{0}^{(i)}$ has been developed through the technique discussed in \cite{26r}.

Consider a fault occurs on the power system between  line $p, q $ at point $r$, where $r$ is the fictitious bus which represents the fault point and $r=n+1$. $m$ denotes the fault distance (in per unit) from bus $p$. Let $Z^{(i)}$ denotes the bus impedance matrix with the addition of fault bus $r$. Ignoring the shunt capacitance we have

\begin{equation}
\label{a1} Z_{kr}^{(i)} = B_{k}^{(i)} + C_{k}^{(i)}m 
\end{equation}

\begin{equation}
\label{a2}
Z_{rr}^{(i)} = A_{0}^{(i)} + A_{1}^{(i)}m + A_{2}^{(i)}m^2 
\end{equation}
Where

\begin{equation} 
B_{k}^{(i)} =Z_{0,pk}^{(i)}   
\end{equation}

\begin{equation} 
C_{k}^{(i)} =Z_{0,qk}^{(i)}-Z_{0,pk}^{(i)}
\end{equation}

Due to the relationship between change in bus voltage and its impedance matrix \cite{24r}, we get

\begin{equation}
\label{a3} 
E_{k}^{(1)} = E_{k}^{(1)0} - Z_{kr}^{(1)}I_{f}^{(1)} 
\end{equation} 

\begin{equation}
\label{a4} 
E_{k}^{(2)} = - Z_{kr}^{(2)}I_{f}^{(2)} 
\end{equation} 

\begin{equation}
\label{a5}
 E_{k}^{(0)} = - Z_{kr}^{(0)}I_{f}^{(0)} 
\end{equation} 
where $E_{k}^{(1)0}$ represents pre-fault positive sequence voltage at bus $k$;
$E_{k}^{(1)}, E_{k}^{(2)}, E_{k}^{(0)}$ are positive, negative and zero sequence voltages during fault at bus $k$ respectively; $I_{f}^{(1)}, I_{f}^{(2)}, I_{f}^{(0)}$ are positive, negative and zero sequence currents during fault at fault point $r$.

Suppose the synchronized positive sequence measurements of pre-fault and fault voltages at two buses $k$ and $l$, which may be away from the faulty transmission line, are available. 

Similar to (\ref{a3}) at bus $l$, we have

\begin{equation}
\label{a6}
E_{l}^{(1)} = E_{l}^{(1)0} - Z_{lr}^{(1)}I_{f}^{(1)} 
\end{equation} 
Eliminating $I_{f}^{(1)} $from (\ref{a3}) and (\ref{a6}), we have

\begin{equation}
\label{a7} 
\frac{{E_{k}^{(1)}-E_{k}^{(1)0}}}{{E_{l}^{(1)}-E_{l}^{(1)0} }} =  \frac{Z_{kr}}{Z_{lr}}  
\end{equation} 
Similar to (\ref{a1}), we get

\begin{equation}
\label{a8}
 Z_{lr}^{(1)} = B_{l}^{(1)} + C_{l}^{(1)}m 
\end{equation}
From (\ref{a1}) and (\ref{a8}), we have

\begin{equation}
\label{a9}
\frac {Z_{kr}}{Z_{lr}} = \frac {B_{k}^{(1)} + C_{k}^{(1)}m }{B_{l}^{(1)} + C_{l}^{(1)}m}  
\end{equation}
From (\ref{a7}) and (\ref{a9}) 

\begin{equation}
\label{a10} 
\frac{{E_{k}^{(1)}-E_{k}^{(1)0}}}{{E_{l}^{(1)}-E_{l}^{(1)0} }} = \frac{B_{k}^{(1)} +C_{k}^{(1)}m }{B_{l}^{(1)} + C_{l}^{(1)}m}  
\end{equation}
Defining 

\begin{equation}
\label{a11} 
D_{kl} = \frac {{E_{k}^{(1)}-E_{k}^{(1)0}}}{{E_{l}^{(1)}-E_{l}^{(1)0}}} = \frac{\Delta E_{k}^{(1)}}{\Delta E_{l}^{(1)}} 
\end{equation}
Where $\Delta E_{k}^{(1)}$ and $\Delta E_{l}^{(1)}$ are positive sequence voltage changes at bus $k$ and $l$.
\\
Solving (\ref{a10}), the fault location has been derived as
\begin{equation}\label{a12} m = \frac {{B_{k}^{(1)}-D_{kl}B_{l}^{(1)}}}{D_{kl}C_{l}^{(1)}-C_{k}^{(1)}}  \end{equation}

\begin{remark}
The above algorithm is feasible if there exists a path which passes through faulty line and does not pass through any bus more than once between two buses $k$ and $l$. Otherwise, the voltage changes between these two buses will be constant, and will be independent of the fault location variable. The feasibility of above method is discussed in detail in \cite{24r}.
\end{remark}

\section{Algorithm Based on Based on Synchronized Sparse Current Measurements (SSCM)}
In this section, fault location algorithm utilizing only the sparse current measurements of the two branches is discussed. The current measurements taken for the fault location algorithm may be away from the faulted line. Similar to previous algorithm, the bus impedance matrix technique is also used here. A fictitious bus is added at the fault point and the impedance of all the buses is expressed as a function of fault location. The fault location is derived based on the relationship between change in branch current due to the fault and the transfer impedance.

Let the impedance between the two branches $k$ and $l$ is $z_{kl}^{(i)}$, where $i$ denote the sequence. Let the fault occur on power transmission line between bus  $p-q $ at point $r$. The positive, negative and zero sequence currents $I_{kl}^{(1)}$, $I_{kl}^{(2)}$, $I_{kl}^{(0)}$ can be calculated based on the following categories \cite{25r}.

\begin{equation}\label{a17} I_{kl}^{(1)}=I_{kl}^{(1)0}-\beta{_{kl}}^{(1)}I_{f}^{(1)}
\end{equation}

\begin{equation}\label{a18} 
I_{kl}^{(2)}=-\beta{_{kl}}^{(2)}I_{f}^{(2)}
\end{equation}

\begin{equation}\label{a19} 
I_{kl}^{(0)}=\beta{_{kl}}^{(0)}I_{f}^{(0)} 
\end{equation}
where 

\begin{equation} \label{a17b} 
\beta_{kl}^{(1)}=B_{kl}^{(1)}+C_{kl}^{(1)}  
\end{equation}
and

\begin{equation}  
B_{kl}^{(1)}=\frac{B_{k}^{(1)}-B_{l}^{(1)}}{z_{kl}^{(1)}} 
\end{equation}

\begin{equation} 
C_{kl}^{(1)}=\frac{C_{k}^{(1)}-C_{l}^{(1)}}{z_{kl}^{(1)}} 
\end{equation}

In this fault location method, the sequence branch currents are formulated with respect to terms $\beta{_{kl}}^{(1)}$ and the fault current through branches. The corresponding terms $\beta{_{kl}}^{(1)}$ can be expressed in terms of unknown fault distance $m$. Suppose that the synchronized branch current phasors between buses $k_1$, $l_1$ and the buses $k_2$, $l_2$ are available from (\ref{a17}), we can find the branch currents as

\begin{equation}\label{a20} I_{k_{1}l_{1}}^{(1)}=I_{k_{1}l_{1}}^{(1)0}-\beta{_{k_{1}l_{1}}}^{(1)}I_{f}^{(1)} \end{equation}

\begin{equation}\label{a21} I_{k_{2}l_{2}}^{(1)}=I_{k_{2}l_{2}}^{(1)0}-\beta{_{k_{2}l_{2}}}^{(1)}I_{f}^{(1)}
\end{equation}\\
Eliminating  $I_{f}^{(1)}$ 

\begin{equation}
\frac{I_{k_{1}l_{1}}^{(1)}-I_{k_{1}l_{1}}^{(1)0}}{I_{k_{2}l_{2}}^{(1)}-I_{k_{2}l_{2}}^{(1)0}} = \frac {\beta{_{k_{1}l_{1}}}^{(1)}}{\beta{_{k_{2}l_{2}}}^{(1)}} \end{equation}\\
Substituting values from (\ref{a20}) and (\ref{a21})

\begin{equation}\label{a23}    
= \frac{B_{k_{1}l_{1}}^{(1)} + C_{k_{1}l_{1}}^{(1)}m }{B_{k_{2}l_{2}}^{(1)} + C_{k_{2}}{l_{2}}^{(1)}m}  
\end{equation}

\begin{equation}\label{a24} 
D_{k_{1}l_{1}k_{2}l_{2}} = \frac{I_{k_{1}l_{1}}^{(1)}-I_{k_{1}l_{1}}^{(1)0}}{I_{k_{2}l_{2}}^{(1)}-I_{k_{2}l_{2}}^{(1)0}} 
\end{equation}
Solving (\ref{a23}) and (\ref{a24}), we get 

\begin{equation}\label{a25} 
m = \frac{{B_{k_{1}l_{1}}^{(1)}-D_{k_{1}l_{1}k_{2}l_{2}}B_{k_{2}{l_{2}}}^{(1)}}}{D_{k_{1}l_{1}k_{2}l_{2}}C_{k_{2}l_{2}}^{(1)}-C_{k_{1}l_{1}}^{(1)}} 
\end{equation}

\begin{remark}
This algorithm is applicable only when the current change due to fault between two branches are linearly independent.
\end{remark}

\section{Proposed Fault Location Algorithm Based on Hybrid Voltage and Current Measurements}
The fault location algorithm utilizing more information about power system has superior performance\cite{27r}. For example, two-end algorithms utilize input signals from both terminals of the line thus providing more information to the relay; therefore, its performance is superior than one-end algorithm as discussed in Section I. 

The fault location algorithm using sparse voltage measurements utilize voltage phasors of only two buses of the power system. In order to have more accuracy, it is needed to utilize voltage phasors of more buses and PMUs are needed on buses at important locations in the power system. The fault location algorithm using sparse current measurements utilize current phasors of the two branches of the power system thus having more information about the power system. But the current transformers may get saturated under certain fault conditions of the transmission lines thus compromising the accuracy of fault location algorithm. 

In the proposed fault location algorithm, both SSVM and SSCM have been utilized. The algorithm utilizes the bus voltages of the faulty transmission line or the adjacent buses and the currents of branches other than the faulty transmission line to develop the fault location index.

Similar to previous algorithms, the bus impedance matrix technique is also used here. A fictitious bus is added at fault point. All the bus impedances have been expressed as a function of fault location. Due to the relationship between change in branch current, change in bus voltage to the fault and the fault impedance, the fault location can be estimated.

Suppose that the synchronized current phasors through branch between bus $k_{1}$ and $k_{2}$, and the voltage phasor of bus $l_{1}$ are available. \\
From (\ref{a20}) and (\ref{a6}), we have

\begin{equation}\label{a6a} E_{l_{2}}^{(1)} = E_{l_{2}}^{(1)0} - Z_{l_{2}r}^{(1)}I_{f}^{(1)} \end{equation} 

\begin{equation}\label{a20a} I_{k_{1}l_{1}}^{(1)}=I_{k_{1}l_{1}}^{(1)0}-\beta{_{k_{1}l_{1}}}^{(1)}I_{f}^{(1)}\end{equation}
\\
Eliminating  $I_{f}^{(1)}$ 

\begin{equation} \label{a29b}
\frac {I_{k_{1}l_{1}}^{(1)}-I_{k_{1}l_{1}}^{(1)0}}{{E_{l_{2}}^{(1)}-E_{l_{2}}^{(1)0} }} = \frac {\beta{_{k_{1}l_{1}}}^{(1)}}{Z_{l_{2}r}} \end{equation}

\begin{equation}\label{a23a}    = \frac {B_{k_{1}l_{1}}^{(1)} + C_{k_{1}l_{1}}^{(1)}m }{B_{l_{2}}^{(1)} + C_{l_{2}}^{(1)}m}  \end{equation}
\\
Defining
\begin{equation}\label{a22} D_{k_{1}l_{1}l_{2}} = \frac {I_{k_{1}l_{1}}^{(1)}-I_{k_{1}l_{1}}^{(1)0}}{{E_{l_{2}}^{(1)}-E_{l_{2}}^{(1)0} }} \end {equation}
\\
Solving we get 
\begin{equation}\label{a23b} m = \frac {{B_{k_{1}l_{1}}^{(1)}-D_{k_{1}l_{1}l_{2}}B_{l_{2}}^{(1)}}}{D_{k_{1}l_{1}l_{2}}C_{l_{2}}^{(1)}-C_{k_{1}}{l_{1}}^{(1)}}  \end{equation}

The fault location $m$ can be obtained by solving (\ref{a23b}) as,
Let

\begin{equation}
	\label{s23a}
	B_k^{(1)} = B_{kr} + jB_{kj}
\end{equation}
\begin{equation}
\label{s23b}
C_k^{(1)} = C_{kr} + jC_{kj}
\end{equation}
\begin{equation}
\label{s23c}
B_l^{(1)} = B_{lr} + jB_{lj}
\end{equation}
\begin{equation}
\label{s23d}
C_l^{(1)} = C_{lr} + jC_{lj}
\end{equation}

It can be derived from (\ref{a23b}) that:

\begin{equation}
\label{s23e}
c_2 m^2 + c_1m + c_0 = 0
\end{equation}
Where

\begin{equation}
\label{s23f}
c_2 = C_{kr}^2 + C_{kj}^2 - |d_{kl}|^2 {C_{lr}^2 + C_{lj}^2}
\end{equation}

\begin{equation}
\label{s23g}
c_1 = 2[B_{kr}C_{kr} + B_{kj}C_{kj} - |d_{kl}|^2 (B_{lr}C_{lr} + B_{lj}C_{lj})]
\end{equation}

\begin{equation}
\label{s23h}
c_1 = B_{kr}^2 + B_{kj}^2 - |d_{kl}|^2 (B_{lr}^2 + B_{lj}^2)
\end{equation}
Solving (\ref{s23e}) results in:

\begin{equation}
\label{s23i}
m_1 = \frac{\left( -c_1 + \sqrt{c_1^2 - 4c_2c_0}  \right)}{2c_2}
\end{equation}

\begin{equation}
\label{s23j}
m_2 = \frac{\left( -c_1 - \sqrt{c_1^2 - 4c_2c_0}  \right)}{2c_2}
\end{equation}
The above solution gives fault location through the root whose value is between $0$ and $+1$.

\begin{algorithm}
\caption{Proposed Fault Location algorithm}
\label{algo1}
\begin{itemize}
	\item After fault occurs take the available synchronized current phasors through branch between bus $k_{1}$ and $k_{2}$, and voltage phasor at bus $l_{1}$
	\item Find $ D_{k_{1}l_{1}l_{2}}$ using (\ref{a22})
	\item Find $ m $ using (\ref{a23b})
\end{itemize}
\end{algorithm}

\begin{remark}
The above algorithm is feasible if there exist a path which passes through faulty line and does not passes through any bus more than once between two buses $k$ and $l$. Also this algorithm is applicable only when the current change due to fault between two branches are linearly independent.
\end{remark}

\section{Simulation Case Studies of the Proposed Algorithm}
\begin{figure*}
	\centering
	\includegraphics[width=0.7\textwidth]{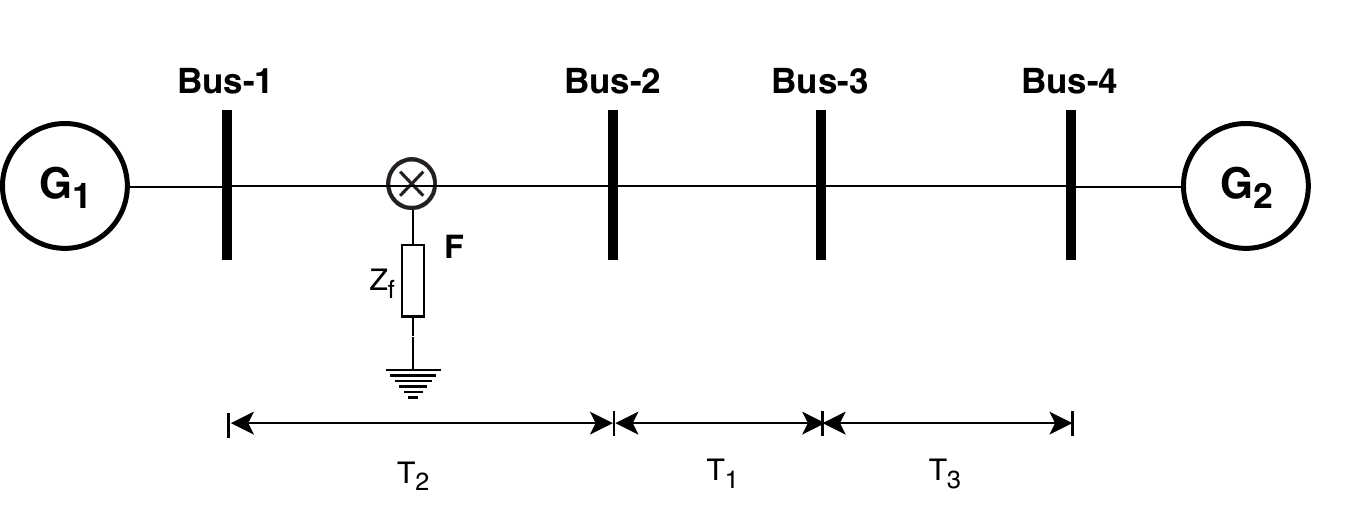}
	\caption{Two-area 4-bus 230 kV, 100 MVA, 50 Hz Power System}
	\label{BLOCKPIC}
\end{figure*}

To evaluate the performance of the proposed fault location algorithm, simulation studies on various test systems are performed. The algorithms are tested on two-area four-bus power system and on IEEE 14-bus system. The tested two-area four-bus power system is a 230 kV, 100 MVA, 50 Hz transmission line system. The block diagram of this power system has been shown in Fig. \ref{BLOCKPIC} and the transmission line and generator data for the system is given in Table \ref{table:1} and \ref{table:2} respectively. The tested four-bus power system and IEEE 14-bus system are simulated in Simulink, while the fault location algorithm is implemented in MATLAB.

\begin{table}
	\caption{Transmission Line Data for Two-area 4-bus System}
	\centering
	\begin{tabular}{|m{4em} | m{3.5em}| m{9em} |m{9em} |}  
		\hline
		\textbf{Transmiss-ion Line}&\textbf{Line Length (Km)} & \textbf{Positive Sequence Impedance (pu) per Km}& \textbf{Zero Sequence Impedance (p.u) per Km}\\ 
		\hline \hline
		\textbf{T2}&178.6 & 0.015455+j0.116066 & 0.098871+j0.365188 \\ 
		\hline 
		\textbf{T1}&21.4 & 0.096188+j0.279293 & 0.243156+j0.822918 \\
		\hline
		\textbf{T3}&91.4 & 0.096188+j0.279293 & 0.243156+j0.822918 \\
		\hline
	\end{tabular}
	\label{table:1}
\end{table}

\begin{table}
	\caption{Generator data for Two-area 4-bus System}
	\centering
	\begin{tabular}{|m{9em} | m{9em}| m{9em}|}  
		\hline
		\textbf{Parameters}  & \textbf{Generator 1} & \textbf{Generator 2} \\  
		\hline \hline
		\textbf{R (pu)}& 0.0006000 & 0.000900 \\
		\hline
		\textbf{L (pu)}& 0.037343 & 0.05423 \\
		\hline
		\textbf{$\mathbf{V_{p-p}\ (V)}$}& 230e3  & 230e3 \\ \hline
	\end{tabular}
	\label{table:2}
\end{table}

\begin{figure}[]
	\centering
	\includegraphics[width=0.7\textwidth]{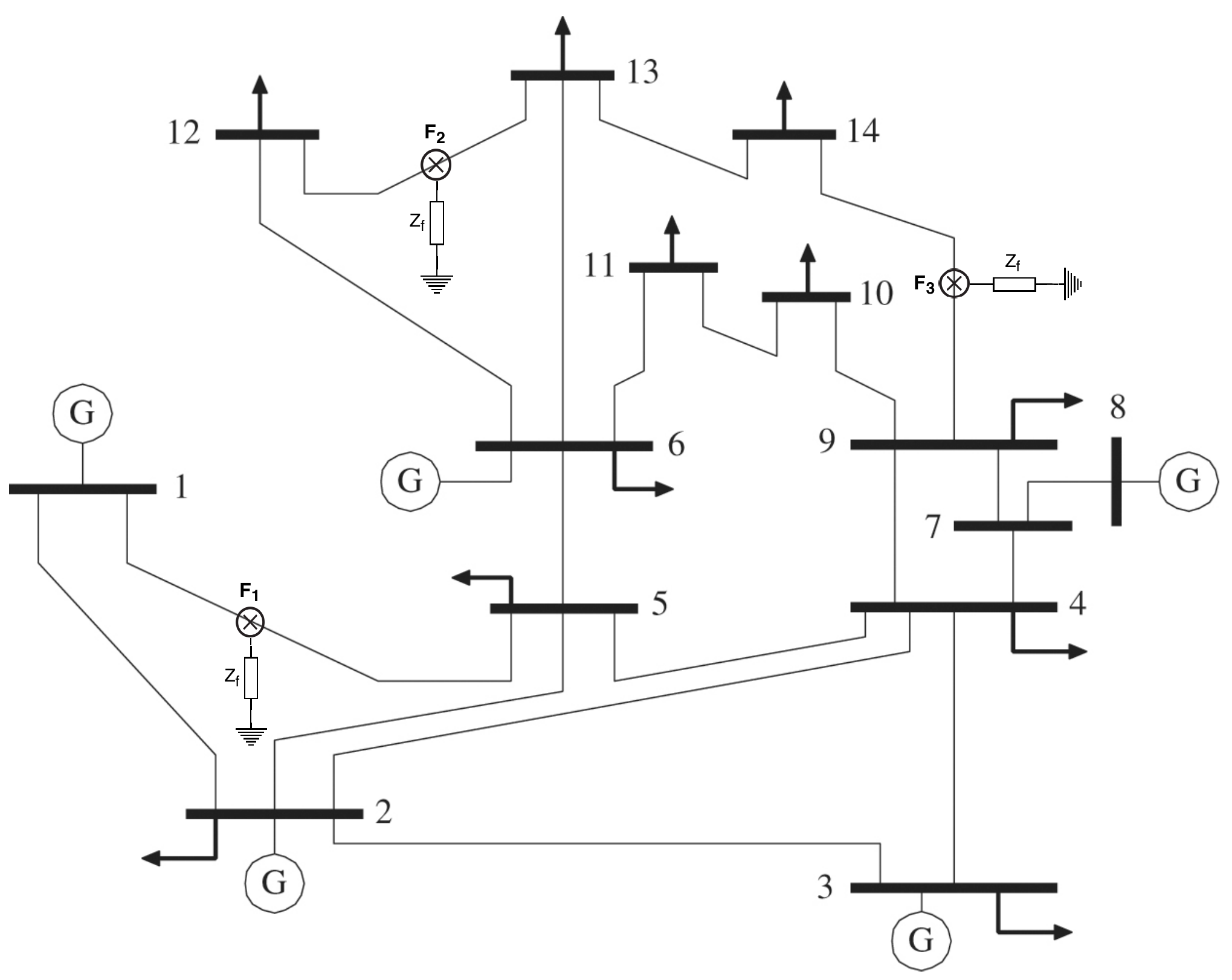}
	\caption{IEEE 14 bus system \cite {ieee}}
	\label{IEEE}
\end{figure}

The IEEE 14-bus system is shown in Fig. \ref{IEEE}. The system data for the 14-bus system is given in \cite{ieee}. PMUs are installed on buses to extract currents and voltage phasors of power system. The transmission lines of the system have been modeled based on lumped parameter model. The shunt capacitance of the lines and the loads is ignored.

The accuracy of fault location algorithm can be evaluated by

\begin{equation}
\label{ac}   
\% \ error= \frac { |Actual \ Location - Estimated \ Location|}{Total \ Length \ of \ Faulted \ Line} \ \times \ 100  
\end{equation} 

\subsection{Case Study of Two-Machine Four-Bus Power System}
In this section, four-bus power system is tested by the proposed fault location algorithm and results are compared with fault location algorithm using SSVM and SSCM under different fault scenarios. The fault is applied on the line $T_2$ at distance $m$ from bus 2. $D$ is the distance from bus 2 to point where fault occurs. Table \ref{table:6} shows fault location results for different fault type i.e. LL, LG, LLL and LLG and for two different values fault resistance i.e. $1~\Omega$ and $10~\Omega$. The percent error can be estimate using (\ref{ac}). Table \ref{table:7} shows calculated percentage error for estimation of fault locations.

\textbf{The results ascertain the accuracy} of the proposed fault location algorithm especially for faults involving two or more than two lines. It is evident from Table \ref{table:7} that the percentage 

\begin{table}
\centering
\caption{Fault Locations Index For Four Bus System}
\begin{tabular}{|m{1.6em}|m{1.5em}|m{1.5em}|m{5.9em}|m{5.9em}|m{5.9em}|} 
\hline
\textbf{Fault Type} & \textbf{D (km)} & $\mathbf{R_{f}}$ ($\mathbf{\Omega}$) & \textbf{Fault location algorithm using SSVM ($V_1$ \& $V_2$)} & \textbf{Fault location algorithm using SSCM ($I_1$ \& $I_2$)} & \textbf{Proposed Fault location algorithm ($V_2$ \& $I_1$)}\\ 
\hline \hline
\multirow{4}{*}{\textbf{LG}}&\multirow{2}{*}{100} & 1 &   102.159215520&99.8788673479 & 99.5493028403    \\ 
\cline{3-6}
 & &10&     102.390658723 &99.35418519 &  100.35351720  \\ 
\cline{2-6}
&\multirow{2}{*}{50}&1 &   51.2426146317& 50.3042991859  &48.6500186505\\ 
\cline{3-6}
 & &10 &   51.3913787694&  49.4995436599 & 49.4878612554 \\ 
\hline
\multirow{4}{*}{\textbf{LL}}& \multirow{2}{*}{100} & 1  &    102.115292583&  99.31826677 &  98.11160195\\ 
\cline{3-6}
  &&10&       102.500170278&9.74874905   &99.78317650\\ 
\cline{2-6}
&\multirow{2}{*}{50}&1 & 51.0167420157&49.2254026616 & 48.4994856406    \\ 
\cline{3-6}
 & &10 &      51.3325513943 &48.7230458688  &49.4878612554  \\ 
\hline 
\multirow{4}{*}{\textbf{LLG}}& \multirow{2}{*}{100} & 1 &  101.709197457 &99.05013477  & 98.26128607   \\ 
\cline{3-6}
  &&10&      102.103928890 &  98.61109167  & 100.04817675\\ 
\cline{2-6}
&\multirow{2}{*}{50}&1 &     51.0152171289  &  49.9819619059&  48.6672773044 \\ 
\cline{3-6}
 & &10 &  51.3129202181& 49.0567192479 & 48.7233159119 \\ 
\hline
\multirow{4}{*}{\textbf{LLL}}& \multirow{2}{*}{100} & 1  &   101.834574675  &98.89487761  &97.91956087   \\ 
\cline{3-6}
  &&10&   102.253251456 &98.50904660   &99.74589643\\ 
\cline{2-6}
&\multirow{2}{*}{50}&1 &   50.9088059273 & 49.0769094421 & 48.4946207551  \\ 
\cline{3-6}
 & &10 &   51.2510792537  &48.6751735675 & 50.9025086341   \\ 
\hline 
\end{tabular}
\label{table:6}
\end{table}

\begin{table}[h!]
\centering
\caption{percent error For Fault Locations Index For Four Bus System }
\begin{tabular}{|m{2em}|m{4.5em}|m{0.8em}|m{5.5em}|m{5.5em}|m{5.5em}|} 
\hline
\textbf{Fault Type} & \textbf{Normalized Distance}  & $\mathbf{R_{f}}$ ($\mathbf{\Omega}$) & \textbf{Fault location algorithm using SSVM ($V_1 \  \& \  V_2$)} & \textbf{Fault location algorithm using SSCM ($I_2$\ \&\ $I_1$)} & \textbf{Proposed Fault location algorithm ($V_2$ \&\ $I_1$)}\\
\hline \hline
\multirow{4}{*}{\textbf{LG}}&\multirow{2}{*}{1.0} & 1 &1.20896726 &  0.06782342&0.2523499   \\ 
\cline{3-6}
 & &10& 1.33855472&  0.36159840&0.1979379  \\ 
\cline{2-6}
&\multirow{2}{*}{0.5}&1 & 0.695752875& 0.17038070&0.75586858  \\ 
\cline{3-6}
 & &10 & 0.77904746 &0.2802101& 0.2868   \\ 
\hline
\multirow{4}{*}{\textbf{LL}}&\multirow{2}{*}{1.0} & 1  & 1.18437434&0.38170952&1.05733373  \\ 
\cline{3-6}
  &&10&   1.39987138&0.70058839&0.12140168  \\ 
\cline{2-6}
&\multirow{2}{*}{0.5}&1 &  0.56928444&0.43370502&0.8401535 \\ 
\cline{3-6}
 & &10 &  0.74610940&0.714979878& 0.2867517   \\ 
\hline
\multirow{4}{*}{\textbf{LLG}}&\multirow{2}{*}{1.0} & 1 &0.95699746 &0.53183873& 0.97352398  \\ 
\cline{3-6}
  &&10&  1.17801169&  0.77766414 &0.2697468 \\ 
\cline{2-6}
&\multirow{2}{*}{0.5}&1 &  0.56843065& 0.0100996&0.74620525 \\ 
\cline{3-6}
 & &10 &0.73511770&0.52815267 &0.7148286    \\ 
\hline
\multirow{4}{*}{\textbf{LLL}}&\multirow{2}{*}{1.0} & 1  &1.02719746 &0.61876951& 1.16485952 \\ 
\cline{3-6}
  &&10&  1.26161896&0.83480026&0.1422751  \\ 
\cline{2-6}
&\multirow{2}{*}{0.5}&1 &  0.50884990&0.51684799&0.84287744  \\ 
\cline{3-6}
 & &10 &  0.70049230 &0.74178407 &0.50532400  \\ 
\hline 
\end{tabular}
\label{table:7}
\end{table}

\subsection{Simulation Results for IEEE 14-bus System}
In this section, IEEE 14-bus system is tested by the proposed fault location algorithm and results are compared with sparse voltage measurement under different fault scenarios. Table \ref{table:8} shows fault location results for fault applied on transmission line between Bus 1 and Bus 5 for different fault type i.e LL, LG, LLL and LLG and for different fault resistance. Table \ref{table:9} shows fault location results for fault applied on line between bus 12 and bus 13 for different fault type and for different fault resistance. Table \ref{table:10} show fault location results for fault applied on transmission line between bus 9 and bus 14 for different fault type and for different fault resistance.

\begin{table}
\caption{Fault Location Index For IEEE 14-Bus system}
\centering
\begin{tabular}{|m{2em} | m{1em}| m{8em} |m{8em} |m{4em}|} 
\hline
\textbf{Fault Type} & $\mathbf{R_{f}}$ ($\mathbf{\Omega}$) & \textbf{Fault location algorithm using sparse voltage measurements ($V_1$\ \& \ $V_5$)} & \textbf{Proposed Fault location algorithm ($V_1$, $I_2$ \& $I_3$)} & \textbf{Normalized Location} \\
\hline \hline
\multirow{8}{*}{\textbf{LLL}} & \multirow{4}{*}{10} &0.694008375157204&0.695133265779887 & 0.7  \\ 
\cline{3-5}
& &0.595910622761881&0.597166994423740 & 0.6\\ 
\cline{3-5}
  &  &0.497447085693533&0.497829632817370 & 0.5  \\ 
\cline{3-5}
  &  &0.400352831169806&0.398116331233032& 0.4  \\ 
\cline{2-5}
 & \multirow{4}{*}{1} &0.691879265784494&0.693525555264493& 0.7 \\ 
\cline{3-5}
 &  &0.594898672100585&0.596871138639339 & 0.6  \\ 
\cline{3-5}
 & &0.499484721411502&0.501805261870535 & 0.5  \\ 
\cline{3-5}
  &  &0.403071139185961&0.405789972288654 & 0.4  \\ 
\hline
\multirow{8}{*}{\textbf{LLG}} & \multirow{4}{*}{10} &0.699421767756495&0.702974549542046 & 0.7  \\ 
\cline{3-5}
 &  &0.601013546970641&0.609642691819562 & 0.6 \\ 
\cline{3-5}
 & &0.498365866804728&00.509218303830038 & 0.5 \\ 
\cline{3-5}
  &  &0.395638551982105&0.409125796598953 & 0.4 \\ 
\cline{2-5}
 & \multirow{4}{*}{1} &0.699567976252138&0.702596568519523& 0.7  \\ 
\cline{3-5}
 &  &0.599833992591630&0.604089002543654 & 0.6  \\ 
\cline{3-5}
 & &0.500023647028278&0.505404857554525 & 0.5   \\ 
\cline{3-5}
  &  &0.399555745288522&0.405851445205361 & 0.4  \\ 
\hline
 \multirow{8}{*}{\textbf{LG}} & \multirow{4}{*}{10} &0.715123600409001&0.727299557032675 & 0.7  \\ 
\cline{3-5}
 &  &0.607090806408158&0.623177555847173 & 0.6  \\ 
\cline{3-5}
 & &0.499443880223307&0.519755660855592 & 0.5  \\ 
\cline{3-5}
  &  &0.391816899736264&0.416806209155708 & 0.4  \\ 
\cline{2-5}
 & \multirow{4}{*}{1} &0.706055281438607&0.711519226343900 & 0.7  \\ 
\cline{3-5}
 &  &0.603751628919271&0.611769434508608 & 0.6  \\ 
\cline{3-5}
 & &0.500470268658642&0.510802460453258 & 0.5  \\ 
\cline{3-5}
  &  &0.396921597097168&0.408954957791715 & 0.4  \\ 
\hline
\multirow{8}{*}{\textbf{LL}} & \multirow{4}{*}{10} &0.708503140769832&0.718864491332760 & 0.7  \\ 
\cline{3-5}
 & &0.602429038352668&0.615171928275021 & 0.6  \\
\cline{3-5}
 &  &0.498048026628321&0.513922900906058& 0.5 \\  
\cline{3-5}
  &  &0.393655443316629&0.413573134298800 & 0.4  \\ 
\cline{2-5}
 & \multirow{4}{*}{1} &0.700505116707854&0.703725447443278 & 0.7  \\ 
\cline{3-5}
 &  &0.600418015997397&0.604976283364740& 0.6  \\ 
\cline{3-5}
 & &0.500066023074132&0.505854509531724 & 0.5  \\ 
\cline{3-5}
  &  &0.399125254154095&0.405905212703340 & 0.4  \\ 
\hline
\end{tabular}
\label{table:8}
\end{table}

\begin{table}[h!]
\caption{Fault Location Index For IEEE 14 Bus System}
\centering
\begin{tabular}{  |m{2em} | m{1em}| m{8em} |m{8em} |m{4em}| } 
\hline
\textbf{Fault Type} & $\mathbf{R_{f}}$ ($\mathbf{\Omega}$) & \textbf{Fault location algorithm using sparse voltage measurements ($V_{12}$ \& $V_{13}$)} & \textbf{Proposed Fault location algorithm ($V_{12}$, $I_{13}$ \& $I_{14}$)} & \textbf{Normalized Location}\\
\hline \hline
\multirow{7}{*}{\textbf{LG}} & \multirow{4}{*}{0.1} &0.373394016982284&0.308250760959466 & 0.3  \\ 
\cline{3-5}
 &  &0.457441403079066&0.393358707854127 & 0.4  \\ 
\cline{3-5}
 & &0.537420126865468&0.481838436558263 & 0.5  \\ 
\cline{3-5}
  &  &0.617998209552984&0.562153545050291 & 0.6  \\ 
\cline{2-5}
 & \multirow{3}{*}{1} &0.426517050685618&0.395988470543525& 0.4 \\ 
\cline{3-5}
 &  &0.514905765469384&0.487337833844390 & 0.5  \\ 
\cline{3-5}
 & &0.606587631246377&0.585044732587657 & 0.6  \\ 
\hline
\multirow{7}{*}{\textbf{LLG}} & \multirow{4}{*}{0.1} &0.366067589608033&0.302047196584776& 0.3  \\ 
\cline{3-5}
 &  &0.448269022723902&0.386796484341236 & 0.4  \\ 
\cline{3-5}
 & &0.531943437105445&0.474040422026917 & 0.5  \\ 
\cline{3-5}
  &  &0.615768651552966&0.562587172213245 & 0.6 \\ 
\cline{2-5}
 & \multirow{3}{*}{1} &0.389305709882585&0.375997865390389& 0.4 \\ 
\cline{3-5}
 &  &0.493469557448198&0.487283788647655 & 0.5  \\ 
\cline{3-5}
 & &0.597970484311008&0.603712594875921& 0.6  \\ 
\hline
\multirow{4}{*}{\textbf{LLL}} & \multirow{4}{*}{0.1} &0.363568021149276&0.298514580918401 & 0.3 \\ 
\cline{3-5}
 &  &0.450485371772019&0.387028167662110 & 0.4 \\ 
\cline{3-5}
 & &0.531817677598756&0.471306209798078 & 0.5   \\ 
\cline{3-5}
  &  &0.618864089620345&0.562319810756823 & 0.6  \\ 
\hline 
\multirow{7}{*}{\textbf{LL}} & \multirow{4}{*}{0.1} &0.363831131706330&0.294169188433461 & 0.3  \\ 
\cline{3-5}
 &  &0.447188403073161&0.380296670785843 & 0.4 \\ 
\cline{3-5}
 & &0.531106401894103&0.467204381361077 & 0.5  \\ 
\cline{3-5}
  &  &0.615756746286453&0.557798630947307 & 0.6  \\ 
\cline{2-5}
 &  \multirow{3}{*}{1} &0.370172664878618&0.408742960327051 & 0.4  \\ 
\cline{3-5}
 & &0.506036757650159&0.477720159271054 & 0.5  \\ 
\cline{3-5}
  &  &0.603709221701594&0.590856707863988 & 0.6  \\ 
\hline
\end{tabular}
\label{table:9}
\end{table}

\begin{table}[h!]
\caption{Fault Location Index For IEEE 14 Bus System}
\begin{center}
\begin{tabular}{  |m{2em} | m{1em}| m{8em} |m{8em} |m{4em}| } 
\hline
\textbf{Fault Type} & $\mathbf{R_{f}}$ ($\mathbf{\Omega}$) & \textbf{Fault location algorithm using sparse voltage measurements ($V_9$ \& $V_{14}$)} & \textbf{Proposed Fault location algorithm ($V_9$, $I_{13}$ \& $I_{14}$)} & \textbf{Normalized Location} \\
\hline \hline
\multirow{6}{*}{\textbf{LL}} & \multirow{3}{*}{1} &0.214714217752116&0.232493856761451& 0.3  \\ 
\cline{3-5}
 &  &0.294258558559854&0.338562501766820 & 0.4  \\ 
\cline{3-5}
 & &0.372845250782083&0.417357990323245& 0.5   \\ 
\cline{2-5}
 & \multirow{3}{*}{10} &0.219151535144925&0.252495885971263 & 0.3  \\ 
\cline{3-5}
 &  &0.305030514224188&0.339299558598834 & 0.4  \\ 
\cline{3-5}
 & &0.390165366103955&0.425997556950395 & 0.5   \\ 
\hline
\multirow{3}{*}{\textbf{LLL}} & \multirow{3}{*}{1} &0.211008613962196&0.265559639603431 & 0.3  \\ 
\cline{3-5}
 &  &0.290243295208899&0.342958492741674 & 0.4  \\ 
\cline{3-5}
 & &0.356868809143634&0.408492563614304 & 0.5  \\ 
\hline
\multirow{6}{*}{\textbf{LLG}} & \multirow{3}{*}{1} &0.217197053015575&0.264751186706570 & 0.3  \\ 
\cline{3-5}
 &  &0.298494681756693&0.345041060204805& 0.4 \\ 
\cline{3-5}
 & &0.379658214843640&0.427034600724855& 0.5 \\ 
\cline{2-5}
 & \multirow{3}{*}{10} &0.220616432614374&0.254160062101079 & 0.3  \\ 
\cline{3-5}
 &  &0.309790390945167&0.345050939358163 & 0.4  \\ 
\cline{3-5}
 & &0.397871273120087&0.435293844249237 & 0.5  \\ 
\hline
\multirow{9}{*}{\textbf{LG}} & \multirow{3}{*}{1} &0.216987211996862&0.252188846473446 & 0.3 \\ 
\cline{3-5}
 &  &0.299161662792965&0.332658085558239 & 0.4  \\ 
\cline{3-5}
 & &0.383636146082595&0.420116993663723 & 0.5  \\ 
\cline{2-5}
  & \multirow{3}{*}{10}  &0.213724205176885&0.258921442732650 & 0.3  \\ 
\cline{3-5}
 &  &0.290893603417413&0.333610843140300& 0.4\\ 
\cline{3-5}
 &  &0.369039687098750&0.411022673765477 & 0.5\\ 
\cline{2-5}
 & \multirow{3}{*}{0.1} &0.215918888869866&0.249108481959334 & 0.3\\ 
\cline{3-5}
  &  &0.297903006928257&0.331456164512131 & 0.4\\ 
\cline{3-5}
 &  &0.384711219463147&0.422993170037254 & 0.5\\ 
\hline
\end{tabular}
\label{table:10}
\end{center}
\end{table}

\subsection{Comparison between Three Fault Location Algorithms}
The proposed fault location algorithm is tested for different fault locations, fault type and fault resistances. Quite satisfactory results were obtained using the proposed algorithm. The fault location estimate depends upon fault type and fault resistance. 
The fault location algorithm using SSVM is completely immune to CT saturation but it can be seen from Table \ref{table:7} that fault location algorithm using SSCM gives relatively accurate estimate than the SSVM. The reason is that it uses more information about the system. Though, the disadvantage lies with the algorithm is that the current transformer may get saturated in case if the available branches current measurements contain branch current of faulty line. 

In order to have more accurate estimates by using sparse voltage measurement, we need to utilize the bus voltages of more number of buses but, in practice, it is uneconomical to have large number of PMUs installed on power system. The proposed fault location algorithm is more accurate than the sparse voltage measurement algorithm as shown in Table \ref{table:7}, \ref{table:8}, \ref{table:9} and \ref{table:10}. The proposed algorithm is also immune to the CT saturation because it uses the bus voltage of faulty line or close to faulty line and branch current of line other than the faulty line. As it uses branch current of line other than faulty line so CT saturation does not occur in this case.


\section{Conclusion}
In this paper, a hybrid fault location algorithm based on synchronized sparse voltage and current measurements has been presented. The proposed method addresses the performance limitation of fault location algorithms based on only voltage or current measurements. The input signals to proposed algorithm are synchronized bus voltage phasors of the faulty line or the bus close to the faulty line. Additionally, synchronized branch current phasors of the line other than faulty line are also input to the algorithm to enhance its accuracy. Simulations on two-area four-bus system and IEEE 14 bus distribution network have been performed to the test the validity of the algorithm. Various types of faults at different locations with varying fault resistances have been applied to the system. Simulation results show that proposed fault location algorithm is robust and offers more accuracy than the algorithms based on only sparse voltage measurements and also it is immune to CT saturation.


%

%

%
%

\ifCLASSOPTIONcaptionsoff
  \newpage
\fi


\begin{thebibliography}{2}
	

\bibitem{1r} 
Crossley PA, and Southern E, The impact of the global positioning system (GPS) on protection \& control,
\textit{ In: Proc of 11th Int Conf on Power System ProtectionPSP,} Bled,pp 15, 1998.

\bibitem{2r} 
Fardanesh B, Zelingher S, and Meliopolos, Multifunctional synchronized measurement network,
\textit{ IEEE Comput Appl in Power,} 11(1):2630, 1998.


\bibitem{3r}
Izykowski J, Rosolowski E, and Saha MM, 
Locating faults in parallel transmission lines under availability of complete measurements at one end, 
\textit{IEE Proc  Gener Transm Distrib,}
151(2):268273, 2004.
\bibitem{4r}
Kawady T, and Stenzel J, 
A practical fault location approach for double circuit transmission lines using single end data,
\textit{IEEE Trans on Power Deliv,}18(4):11661173, 2004.
\bibitem{5r}
Takagi T, Yamakosi Y, and Yamura M,
Development of new type fault locator using the oneterminal voltage and current data,
\textit{IEEE Trans on PAS,} 101(8):28922898, 1981.
\bibitem{6r}
Zhang Q, Zhang Y, and Song W,
Transmission line fault location for phase-to-earth fault using one-terminal data,
 \textit{IEE Proc  Gener Transm Distrib,} 146(2):121124, 1999.
\bibitem{7r}
Zhang Q, Zhang Y, and Song W,
Fault location of two-parallel transmission line for non-earth fault using one-terminal data,
\textit{IEEE Trans on Power Deliv,}14(3):863867, 1999.
\bibitem{8r}
Zhang Y, Zhang Q, and Song W, 
Transmission line fault location for double phaseto-earth fault on non-direct-ground neutral system,
\textit{IEEE Trans on Power Deliv,}15(2):520524, 1999.
\bibitem{29r}
Eriksson L, Saha MM, and Rockefeller GD,  
An accurate fault locator with compensation for apparent reactance in the fault resistance resulting from remote end infeed,
\textit{IEEE Trans on PAS,} 104(2):424–436,1985.


\bibitem{9r} 
Jiang J-A, Lin Y-H, and Yang J-Z, 
An adaptive PMU based fault detection/location technique for transmission lines - Part II: PMU implementation and performance evaluation,
\textit{ IEEE Trans on Power Deliv,} 15(4):11361146, 2000.
\bibitem{10r} 
Jiang J-A, Lin Y-H, and Yang J-Z,
An adaptive PMU based fault detection/location technique for transmission lines - Part I: Theory and algorithms,
\textit{ IEEE Trans on Power Deliv,} 15(2):486493, 2000.
\bibitem{11r}
Johns AT, and Jamali S,
Accurate fault location technique for power transmission lines, 
\textit{ IEE Proc C,} 137(6):395402, 1990.
\bibitem{12r}
Kezunovic M, Mrkic J, and Perunicic B,
An accurate fault location algorithm using synchronized sampling,
\textit{ Electr Power Syst,}Res 29(3):161169, 1994.
\bibitem{13r}
Kezunovic M, and Perunicic B, 
Automated transmission line fault analysis using synchronized sampling at two ends,
\textit{ IEEE Trans on Power Syst,} 11(1):441447, 1996.
\bibitem{14r}
Lin Y-H, Liu C-W, and Chen C-S, 
A new PMU-based fault detection/location technique for transmission lines with consideration of arcing fault discrimination-part I: theory and algorithms, 
\textit{IEEE Trans on Power Deliv,}19(4):15871593, 2004.
\bibitem{15r}
Lin Y-H, Liu C-W, and Yu C-S,
A new fault locator for three-terminal transmission linesusing two-terminal synchronized voltage and current phasors,
\textit{IEEE Trans on Power Deliv,}17(2):452459, 2002.
\bibitem{16r}
Girgis AA, Hart DG, and Peterson WL,
A new fault location technique for two- and three-terminal lines,
\textit{ IEEE Trans on Power Deliv,} 7(1):98–107, 1992.
\bibitem{17r}
Kezunovic M, and Perunicic B,
Automated transmission line fault analysis using synchronized sampling at two ends,
\textit{IEEE Trans on Power Syst,} 11(1):441–447, 1996.


\bibitem{18r} 
M.B. Djuri6, Z.M. Radojevi6 and V.V. Terzija, 
Distance Protection and Fault Location Utilizing Only Phase Current Phasors, 
\textit{IEEE Transactions on Power Delivery,} Vol. 13, No. 4, October 1998.
\bibitem{19r}
Brahma SM 
Fault location scheme for a multi-terminal transmission line using synchronized voltage measurements,\textit{ IEEE Trans on Power Deliv,} 20(2):1325–1331, 2005. 
\bibitem{20r}
Brahma SM, and Girgis AA, 
Fault location on a transmission line using synchronized voltage measurements,
\textit{IEEE Trans on Power Deliv,}19(4):1619–1622, 2004.
\bibitem{21r}
Z. I, M. JF, and M. AJ, 
Fault location on two-terminal transmis-sion lines based on voltages, in \textit{IEE Proc Gener Transm Distrib,}vol. 143, pp. 1-6, 1996.

\bibitem{22r}
Tziouvaras DA, Roberts J, Benmouyal G,
New multi-ended fault location design for two- or three-terminal lines,
\textit{ Proc of 7th Int Conf on Developments in Power System
Protection  DPSP, IEE CP476} pp 395398, 2001.
\bibitem{23r}
Tziouvaras DA, Roberts J, Benmouyal G,
New multi-ended fault location design for two- or three-terminal lines.
\textit{ Proc of CIGRE  Study Committee 34 Colloquium and
Meeting, Florence,} paper 213,1999.



\bibitem{24r} 
Yuan Liao,
Fault Location for Single-Circuit Line Based on Bus-Impedance Matrix Utilizing Voltage Measurements,
\textit{IEEE TRANSACTIONS ON POWER DELIVERY,} VOL. 23, NO. 2, APRIL 2008.

\bibitem{ietgtd}
A. Ali, A. Q. Khan, B. Hussain, M. T. Raza and M. Arif, "Fault modelling and detection in power generation, transmission and distribution systems," in \textit{IET Generation, Transmission \& Distribution}, vol. 9, no. 16, pp. 2782-2791, 3 12 2015

\bibitem{ifac}
Abdul Qayyum Khan, Qudrat Ullah, Muhammad Sarwar, Sufi Tabassum Gul, Naeem Iqbal,
Transmission Line Fault Detection and Identification in an Interconnected Power Network using Phasor Measurement Units,
\textit{IFAC-PapersOnLine}, Volume 51, Issue 24, 2018, Pages 1356-1363, ISSN 2405-8963


\bibitem{25r} 
Y. Liao, 
Fault location using sparse current measurements,
In \textit{NAPS 2007,} Las Cruces, New Mexico, USA, September 2007.

\bibitem{26r}
K. Takahashi, J. Fagan, and M. Chen, 
Formation of a sparse bus impedance matrix and its application to short circuit study,
in \textit{Proc.8th PICA Conf.}, Minneapolis, MN, Jun. 4–6, 1973, pp. 63–69.


\bibitem{27r}
Murari Mohan Saha ,Jan Izykowski, and Eugeniusz Rosolowski,
Fault Location on Power Networks, 
\textit{Springer-Verlag,} London Limited 2010.
\bibitem{ieee}
JUN ZHU,
Analysis of Transmission System Faults in the Phase Domain, 
B.S, Shanghai Jiaotong University 2004.


\end{thebibliography}
\end{document}